\documentclass[fleqn,usenatbib]{mnras}

\usepackage{newtxtext}
\usepackage[T1]{fontenc}

\DeclareRobustCommand{\VAN}[3]{#2}
\let\VANthebibliography\thebibliography
\def\thebibliography{\DeclareRobustCommand{\VAN}[3]{##3}\VANthebibliography}

\usepackage{graphicx}	% Including figure files
\usepackage{amsmath}	% Advanced maths commands
\usepackage{amssymb}	% Extra maths symbols

\title[HI spin temperature at cosmological distances]{A statistical measurement of the \mbox{H\,{\sc i}} spin temperature in DLAs at cosmological distances}

\author[J. R. Allison]{
James R. Allison$^{1,2}$\thanks{E-mail: james.allison@physics.ox.ac.uk} 
\\
$^{1}$Sub-Dept. of Astrophysics, Department of Physics, University of Oxford, Denys Wilkinson Building, Keble Rd., Oxford, OX1 3RH, UK\\
$^{2}$ARC Centre of Excellence for All-Sky Astrophysics in 3 Dimensions (ASTRO 3D)\\
}

\date{Accepted XXX. Received YYY; in original form ZZZ}

\pubyear{2020}

\begin{document}
\label{firstpage}
\pagerange{\pageref{firstpage}--\pageref{lastpage}}
\maketitle

\begin{abstract} 
Evolution of the cosmic star formation rate (SFR) and molecular gas mass density is expected to be matched by a similarly strong evolution of the fraction of atomic hydrogen (\mbox{H\,{\sc i}}) in the cold neutral medium (CNM). We use results from a recent commissioning survey for intervening 21-cm absorbers with the Australian Square Kilometre Array Pathfinder (ASKAP) to construct a Bayesian statistical model of the $N_{\rm HI}$-weighted harmonic mean spin temperature ($T_{\rm s}$) at redshifts between $z = 0.37$ and $1.0$. We find that $T_{\rm s} \leq 274$\,K with 95\,per\,cent probability, suggesting that at these redshifts the typical \mbox{H\,{\sc i}} gas in galaxies at equivalent DLA column densities may be colder than the Milky Way interstellar medium ($T_{\rm s, MW} \sim 300$\,K). This result is consistent with an evolving CNM fraction that mirrors the molecular gas towards the SFR peak at $z \sim 2$. We expect that future surveys for \mbox{H\,{\sc i}} 21-cm absorption with the current SKA pathfinder telescopes will provide  constraints on the CNM fraction that are an order of magnitude greater than presented here.
\end{abstract}


%%%%%%%%%%%%%%%%%%%%%%%%%%%%%%%%%%%%%%%%%%%%%%%%%%

%%%%%%%%%%%%%%%%% BODY OF PAPER %%%%%%%%%%%%%%%%%%

\section{Introduction}

The coldest ($T_{\rm k} < 100$\,K) interstellar gas has a fundamental role in forming stars and fuelling galaxy evolution throughout cosmic history. Understanding why the star formation rate (SFR) density of the Universe has declined rapidly since peaking at $z \approx 2$ (e.g. \citealt{Hopkins:2006, Madau:2014, Driver:2018}) is intimately tied to determining how the mass fraction of cold gas in galaxies has evolved.

Stars form in the dusty molecular gas of the interstellar medium (ISM) and so it is believed that this phase is most important in fuelling evolution of the SFR density throughout cosmic history. Recent surveys for the tracers of the dense molecular gas, principally CO emission (e.g. \citealt{Decarli:2019, Decarli:2020, Lenkic:2020, Riechers:2020a, Riechers:2020b, Fletcher:2021}) and the far-infrared and mm-wavelength dust continuum (e.g. \citealt{Berta:2013, Scoville:2017, Magnelli:2020}), provide evidence that the mass density evolves strongly and roughly mirrors that of the SFR density (albeit with a slightly delayed peak;  \citealt{Tacconi:2020}). In contrast, observations of the diffuse atomic gas, principally using \mbox{H\,{\sc i}} 21-cm emission (e.g. \citealt{Zwaan:2005b, Braun:2012, Rhee:2018, Jones:2018, Chowdhury:2020}) and Lyman-$\alpha$ absorption (e.g. \citealt{Noterdaeme:2012, Zafar:2013, Crighton:2015, Sanchez-Ramirez:2016, Bird:2017}), reveal a mass density that evolves comparatively slowly and declines by only a factor $\sim 2$ from peak SFR to the present day. Recently, \cite{Walter:2020} showed that this global behaviour could be described by a simple phenomenological model of the gas parametrised by the net infall rate of ionised inter/circumgalactic gas, which replenishes the \mbox{H\,{\sc i}} reservoir, and the conversion of \mbox{H\,{\sc i}} to H$_{2}$. Both these processes have declined by an order magnitude since their peak. 

Evolution of the physical state of the atomic gas in galaxies is as yet unknown and could well deviate from that of the total mass density measured from 21-cm emission and Lyman-$\alpha$ absorption surveys. Observations of the Milky Way ISM reveal a multi-phased diffuse neutral medium that spans two orders of magnitude in temperature and density (e.g. \citealt{Heiles:2003, Murray:2018}). Two distinct stable phases, the denser cold neutral medium (CNM; $T_{\rm k} \sim 100$\,K) and more diffuse warm neutral medium (WNM; $T_{\rm k} \sim 10\,000$\,K), co-exist in a pressure equilibrium that is determined by heating and cooling processes that are dependent on the star formation rate, dust abundance and gas-phase metallicity (\citealt{Wolfire:2003}). Further dynamical processes, such as turbulence and supernova shocks, are thought to generate an unstable third phase at intermediate temperatures (UNM; e.g. \citealt{Murray:2018}). If the presence of CNM is a pre-requisite for molecular cloud and star formation (e.g. \citealt{Krumholz:2009}), then a similarly strong redshift evolution would be expected for the CNM fraction.

The \mbox{H\,{\sc i}} 21-cm absorption line detected in the spectra of background radio sources is an effective tracer of the cold phase atomic gas to large redshifts. The equivalent width is inversely related to the excitation (spin) temperature, thereby providing a means by which the relative mass fractions of distinct thermal phases can be inferred. In the few cases where the \mbox{H\,{\sc i}} column density ($N_{\rm HI}$) can determined independently from either 21-cm emission (e.g. \citealt{Reeves:2016, Borthakur:2016, Gupta:2018}) or Lyman-$\alpha$ absorption (see \citealt{Kanekar:2014a} and references therein; hereafter K14), the $N_{\rm HI}$-weighted harmonic mean of the spin temperature can be measured directly. K14 show that the distribution of spin temperatures for Damped Lyman-$\alpha$ absorbers (DLAs; $N_{\rm HI} > 2 \times 10^{20}$\,cm$^{-2}$) at $z > 2.4$ is statistically different (at 4\,$\sigma$ significance) to those at lower redshifts. However it is observationally challenging and expensive to draw a sufficiently large sample of nearby \mbox{H\,{\sc i}} galaxies and distant Lyman-$\alpha$ absorbers in order to provide strong constraints on the redshift of evolution of the cold phase gas. Recently, \cite{Curran:2017, Curran:2019} showed that the 21-cm absorber population, and hence spin temperature, may evolve with the star formation history Universe. However, they use an evolutionary model for the covering fraction of radio sources that would mimic any perceived evolution in the spin temperature. Model-independent methods are therefore required to verify such a claim. An alternative approach is to infer the spin temperature indirectly from much larger 21-cm absorption-line surveys by comparing the outcomes with the expected detection yield (\citealt{Darling:2011, Allison:2016, Allison:2020, Grasha:2020}). 

In this paper we present a statistical measurement of the $N_{\rm HI}$-weighted harmonic mean spin temperature in DLAs at intermediate cosmological redshifts ($0.37 < z < 1.00$). This uses a Bayesian technique first proposed by \cite{Allison:2016} to infer the \mbox{H\,{\sc i}} spin temperature by comparing the expected and actual detection yields in 21-cm absorption-line surveys. The results presented here are derived from a recent commissioning project with the Australian Square Kilometre Array Pathfinder (ASKAP; \citealt{DeBoer:2009,  Hotan:2021}), which carried out a survey for 21-cm absorption lines in a sample of 53 bright compact radio sources (\citealt{Sadler:2020}; hereafter S20). This is the first demonstration of this method to measure the spin temperature at cosmological distances. In future work we plan to use data from the ASKAP First Large Absorption Survey in HI (FLASH; e.g. \citealt{Allison:2016, Allison:2020}) to measure the harmonic mean spin temperature to greater precision and enable measurements as a function of redshift. 

Throughout this paper we use a flat $\Lambda$CDM cosmology with $\Omega_{\rm m} = 0.3$, $\Omega_{\Lambda} = 0.7$ and $H_{0} = 70$\,km\,s$^{-1}$\,Mpc$^{-1}$ (e.g. \citealt{Spergel:2007}).

\section{Data analysis}

\subsection{Data}

We use data from the 21-cm absorption-line survey carried out by S20 during commissioning of the ASKAP telescope. S20 searched 21-cm redshifts between $z = 0.37$ and $1.0$ towards a sample of 53 radio sources selected from the Australia Telescope 20\,GHz (AT20G) Survey catalogue (\citealt{Murphy:2010}). They detected four intervening \mbox{H\,{\sc i}} 21-cm absorbers towards PKS\,0834$-$20, PKS\,1229$-$02, PKS\,1610$-$77 and PKS\,1830$-$211, as well as 21-cm absorption associated with the radio galaxy PKS\,B1740$-$517 and intervening OH 18-cm absorption towards PKS\,1830$-$211. Although the intervening absorbers towards PKS\,1229$-$02 and PKS\,1830$-$211 were already known, the sample was selected based on source flux density and declination and so is not biased by known detections. We refer the reader to S20 for further details of the sample selection, observations and individual detections (see also \citealt{Allison:2017, Allison:2019} for further details of the ASKAP detections towards PKS\,1830$-$211 and PKS\,1740$-$517).

Each radio source spectrum records the fractional absorption of the background continuum flux density as a function of observed frequency, along with a measurement of the noise per 18.5\,kHz spectral channel. The velocity resolution ranges between 5.3 and 7.8\,km\,s$^{-1}$ and the spatial resolution, using natural weighting, is approximately 1.8\,arcmin for the 6-antenna ASKAP Boolardy Engineering Test Array (BETA; \citealt{Hotan:2014, McConnell:2016}) and 50\,arcsec for the ASKAP-12 array (\citealt{Hotan:2021}). The spectral baseline is flat for the dynamic range requirements of this survey, although as noted by S20 there is an additional non-Gaussian noise component for the ASKAP-12 spectra, which is discussed further in \autoref{section:reliability}. We use these data to determine the sensitivity to the 21-cm optical depth for each spectral interval, thereby allowing us to estimate the expected number of intervening 21-cm absorber detections as a function of the properties of the foreground \mbox{H\,{\sc i}} and background sources. 

\subsection{Absorption line detection}

We automate line detection using a bespoke software tool called \textsc{FLASHfinder}\footnote{\url{https://github.com/drjamesallison/flashfinder}} (\citealt{Allison:2012}), which uses the \textsc{PyMultiNest} (\citealt{Buchner:2014}) implementation of \textsc{MultiNest} (\citealt{Feroz:2008, Feroz:2009, Feroz:2013}). The multi-modal capability of \textsc{MultiNest} enables more than one line to be detected in a given spectrum. Detection significance is given by the Bayes factor $B$ (e.g. \citealt{Kass:1995}), a statistic that is equal to the ratio of Bayesian evidences for a line model to a null model. Since no prior preference is given for either the line or null models, which is reasonable if we are testing for the unknown incidence of 21-cm absorption lines, $B$ is equal to the odds in favour of the line model. To calculate the Bayesian evidence we use the likelihood function for a normal distribution, with standard deviation equal to the measured rms noise in each spectral channel. 

For the purpose of 21-cm absorption line detection we use a single Gaussian profile to model the optical depth in velocity, which is then converted into the observed absorption line. We use non-informative priors for the model parameters within ranges set by the data and physically realistic limits; for the line position we use a uniform prior over the full range of the spectrum, for the full width at half maximum (FWHM) we use a loguniform prior between 0.1 and 2000\,km\,s$^{-1}$, and for the peak optical depth we use a loguniform prior between 1 per cent of the median rms noise per 18.5\,kHz channel and a maximum value of 100. 

A Gaussian profile is appropriate given that broadening is expected to be dominated by Doppler shift due to line-of-sight thermal, turbulent and bulk motion of the gas. A typical 21-cm absorption line will have a more complex velocity profile due to the relative kinematic and spatial distributions of the foreground absorber and background radio source. However, the effect of more complex model choices on detecting high signal-to-noise lines is not expected to be significant. 

\subsection{Detection reliability}

\begin{figure}
\centering
\includegraphics[width=0.45\textwidth]{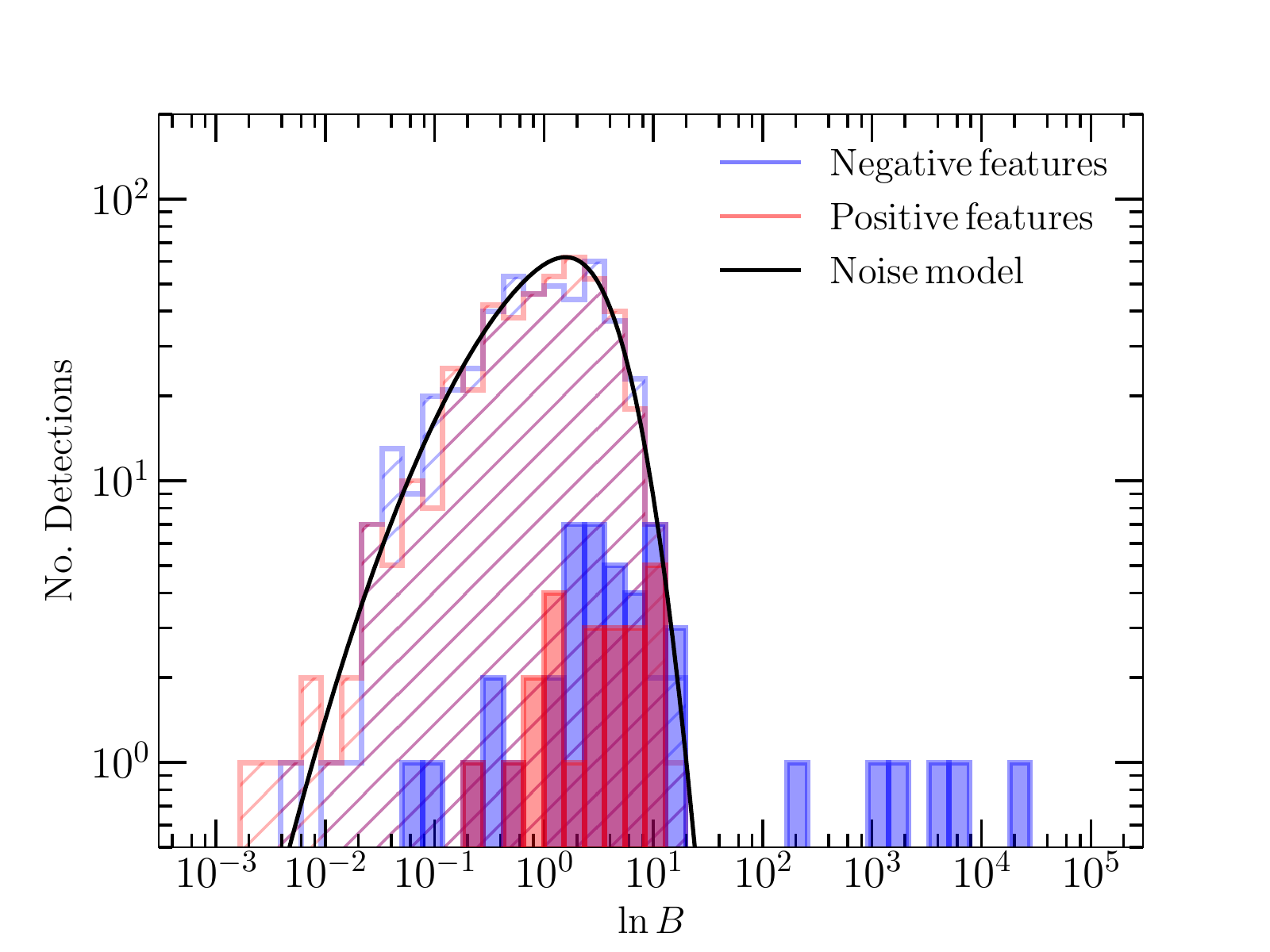}
\caption{The distribution of positive and negative features detected in the data as a function of the detection statistic, the Bayes factor $B$. Solid histograms denote results for the S20 spectra. For comparison, the empty hatched histograms denote the larger data set of the ASKAP wide-field survey of the GAMA\,23 field, undertaken during FLASH early science (A20). Regions of overlapping negative and positive features are coloured magenta. The solid black line denotes a model fit to the distribution of positive features detected in the A20 spectra. There are six negative features that are clearly reliable, of which four are intervening 21-cm absorbers, one is 21-cm absorption associated with the radio galaxy PKS\,B1740$-$517 and one is OH 18-cm absorption towards PKS\,1830$-$211 (see S20 for further details).}
\label{figure:reliability}
\end{figure} 

Our detection statistic, $B$, is the Bayesian odds in favour of a feature represented by the absorption-line model, with features at $\ln{B} \gtrsim 3$ strongly favoured. Artefacts that remain in the data after processing, and that are not accounted for by the null model, will reduce the detection reliability (also known as the purity or fidelity). This issue is particularly problematic for absorption-line surveys, where any bandpass error not corrected by calibration is amplified in the source spectrum. Possible solutions include further data processing techniques to remove or mask artefacts, adapting the null model to account for them, or distinguishing them from true lines in parameter space, \emph{a posteriori}. Here we use the latter method, focusing specifically on the detection statistic $B$, but will explore other approaches in future work.

\subsubsection{Using positive features to determine reliability}\label{section:reliability}

We characterise the distribution of false absorption-like detections by running our detection method on the inverted spectra, and then inspect the distribution of positive (emission-like) features detected as a function of $B$. This is analogous to the procedure in blind surveys for \mbox{H\,{\sc i}} 21-cm emission in the nearby Universe (e.g. \citealt{Serra:2012, Serra:2015}) and CO emission at higher redshifts (e.g \citealt{Walter:2016, Pavesi:2018, Lenkic:2020}) that use negative (absorption-like) features to determine reliability. The implicit assumption is that the incidence of true lines amongst the reference distribution is negligible compared with the incidence of features due to noise and/or artefacts. We discuss the validity of this assumption for our data in \autoref{section:contaminating_lines} below.

In \autoref{figure:reliability} we show the distribution of detected negative and positive features as a function of $B$. For comparison, we include results from the larger ASKAP wide-field blind survey of the GAMA\,23 field by \cite{Allison:2020} (hereafter A20), which was undertaken during FLASH early science. The reliability of both surveys is limited by a non-Gaussian contribution to the spectral noise, at the level of $\sim 1$\,per\,cent of the signal, which is caused by incorrect firmware weights used for correcting the coarse channelisation in the ASKAP-12 array\footnote{This error is not present in the earlier 6-antenna BETA data, and has been corrected for observations with the full ASKAP-36 array.}. Because the additional noise is multiplicative, it is preferentially detected towards sources with higher signal-to-noise continuum. This behaviour is evident in the results shown \autoref{figure:reliability}; the distribution of positive features detected in S20 is skewed to higher $\ln(B)$ values than A20, which is consistent with the proportion of higher signal-to-noise spectra in that sample.

We note that in blind surveys for CO emission the reliability of low signal-to-noise lines is typically determined by modelling the distribution of false detections assuming negative/positive symmetry, thereby enabling fainter sources to be included in the CO luminosity function. However, for the data considered here it is not certain that the additional positive and negative features generated by the channelisation error would be distributed symmetrically. Without further characterisation of the additional noise we cannot determine the reliability of excess negative features within the low signal-to-noise region.  We therefore take a conservative approach by selecting a $\ln(B)$ threshold above which we are confident that any absorption-like features are reliable. 

We capture the behaviour of the well-sampled distribution of positive features in the A20 spectra by fitting a skew normal distribution in the logarithm of $\ln{B}$, with parameter $\lambda = -3$ (\citealt{Azzalini:1996}). We stress that there was no basis for this choice other than the model provides a reasonably accurate analytical representation of the observed distribution. We extrapolate from this model that there is less than 0.003 probability for positive features above $\ln{B} = 16$, corresponding to an expectation of less than one detection. This is the reliability threshold adopted by A20 to determine the completeness of those data. In the case of the S20 spectra, the distributions are relatively under-sampled and harder to model; we also note the excess negative features in the low signal-to-noise regime that could either be caused by real absorption lines or simply bias in the noise behaviour. Although it is not possible to determine if some of these excess negative features are true intervening 21-cm absorbers, none are detected at the 21-cm line position in the spectrum corresponding to the source redshift and so we can rule out associated absorption. Given that the distribution appears to be skewed to higher $\ln{B}$ than that of A20, we choose a threshold of $\ln{B} = 20$, above which no detections of negative or positive features are apparent in the low signal-to-noise distribution.

\subsubsection{Consideration of other spectral lines}\label{section:contaminating_lines}

As previously mentioned, our reliability analysis assumes that the reference distribution of positive (emission-like) features is dominated by noise and not contaminated with a significant fraction of true spectral lines. We discuss the validity of that assumption here. 

First we consider the possibility that the positive features may contain detections of \mbox{H\,{\sc i}} 21-cm emission. We note that none of the positive features are detected at the 21-cm line position in the spectrum corresponding to the source redshift, and so we can rule out emission associated with the host galaxy. However, it is also possible that we may detect \mbox{H\,{\sc i}} emission from other galaxies in our line of sight.  In the case of the S20 spectra, the lowest redshift is $z = 0.37$ with a typical spectral rms noise of  $\sim 10$\,mJy\,beam$^{-1}$ per 18.5\,kHz, which for a line width of $\sim 100$\,km\,s$^{-1}$ gives a 5-sigma detection limit of $M_{\rm HI} \sim 7.8 \times 10^{11}$\,M$_{\odot}$ (e.g. \citealt{Meyer:2017}). Likewise, for the A20 spectra, the lowest \mbox{H\,{\sc i}} redshift is $z = 0.34$ with a median rms noise of 3.2\,mJy\,beam$^{-1}$, giving a corresponding mass limit of $M_{\rm HI} \sim 2.1 \times 10^{11}$\,M$_{\odot}$. Both limits are well beyond the high-mass end of the local \mbox{H\,{\sc i}} mass function (e.g. \citealt{Jones:2018}) and so we conclude that the serendipitous detection of 21-cm emission in these spectra is very unlikely. 

Of the other emission lines that we could possibly detect at these frequencies, the most likely are luminous masing emission from the main 1665 and 1667\,MHz lines of the hydroxyl radical (OH). The S20 spectra cover OH redshifts between $z = 0.61$ and $1.34$, which for a maser line width of $\sim 100$\,km\,s$^{-1}$ corresponds to a 5-sigma detection limit between $L_{\rm 1665} = 0.3$ and $1.8 \times 10^{5}$\,$L_{\odot}$.  In the case of the A20 spectra, they cover OH redshifts between $z = 0.57$ and $1.10$ with a corresponding 5-sigma detection limit between $7.3 \times 10^{3}$ and $3.7 \times 10^{4}$\,$L_{\odot}$. First we consider OH emission (and absorption) associated with host galaxies of both the radio sources and the reliably identified 21-cm absorbers. Only the known OH absorption line towards PKS1830$-$211 is detected. Secondly, we consider the possibility of serendipitously detecting line-of-sight megamaser emission. We can use the low-redshift ($z < 0.23$) OH megamaser luminosity function  measured by \cite{Darling:2002} to estimate the expected number of detections. Assuming a total sky area covered by the S20 spectra $\sim 0.05\,\deg^{2}$ (based on the angular resolution of each spectrum) the expected number of serendipitously detected OH megamasers is negligible (i.e. $\mathcal{N}_{\rm OH} \ll 1$), even if the luminosity function evolves by an order of magnitude at higher redshifts. Likewise, for the 1253 spectra of A20 the total sky area covered is $\sim 0.76\,\deg^{2}$, again giving a negligible number of expected OH megamasers.

Clearly contamination from other emission and absorption lines is not an issue with these data. However, future large 21-cm absorption surveys with the SKA and pathfinder telescopes will be sensitive to a greater volume and will need to take into consideration these other emission and absorption lines when determining the reliability of their detections. For these future surveys, in addition to OH, we will need to consider that other molecular species at high redshift (for example the H$_{2}$CO 6-cm line) could give rise to confusing absorption lines that would affect 21-cm absorption reliability. Furthermore, there are other masing transitions that lead to positive features that would contaminate the reference distribution, such as the radio-frequency recombination lines and the conjugate OH satellite lines at 1612 and 1720\,MHz.

\subsection{Completeness}\label{section:completeness}

\begin{figure}
\begin{center}
\includegraphics[width=\columnwidth]{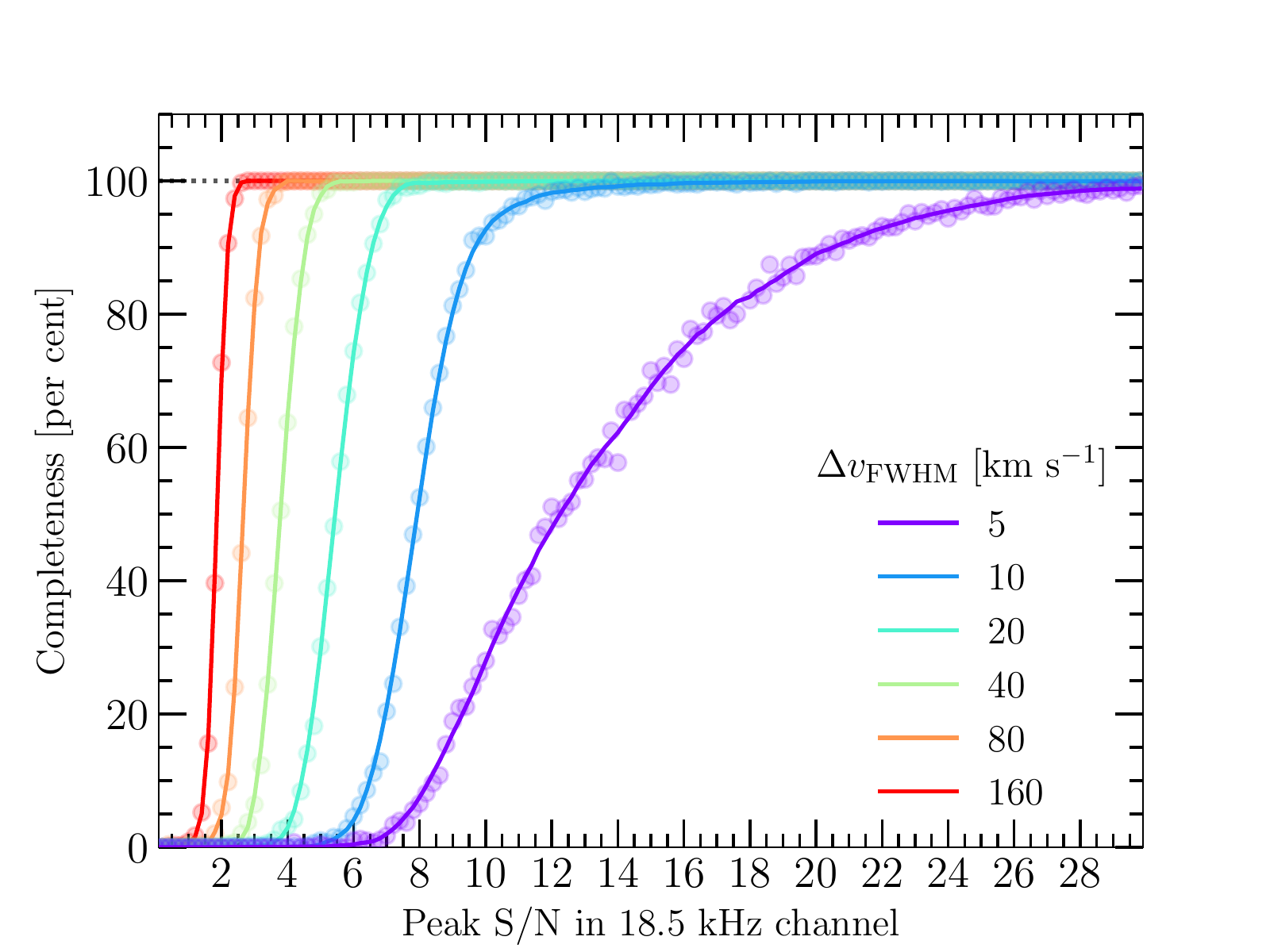}
\caption{The expected completeness of reliable detections ($\ln(B) > 20$) in the S20 spectra, as a function of the peak S/N in a single 18.5-kHz (5.3 -- 7.8\,km\,s$^{-1}$) channel. The lines denote smoothing of the simulated data (circles) using a Savitzky-Golay filter (\citealt{Savitzky:1964}).}\label{figure:completeness}
\end{center}
\end{figure}
 
We define the completeness as the probability that an absorption line with a given peak signal-to-noise ratio (S/N) and width is recovered from the data using our detection method and reliability threshold. This is determined by randomly populating our sample spectra with 1000 Gaussian-profile absorption lines in bins of peak S/N and FWHM and measuring the fraction that are recovered. The results are shown in \autoref{figure:completeness} and define the completeness as a bivariate function of the peak S/N and FWHM. 

We now consider how these are used to determine the completeness in terms of the physical properties of absorption lines. For a line with peak optical depth $\tau$, the peak S/N in $i$'th spectral channel towards the $j$'th source is given by
\begin{equation}\label{equation:signal_to_noise}
    \mathrm{(S/N)}_{i,j} = \left[{1 - \exp(-\tau)}\right] \frac{c_{\rm f}\,S^{\rm c}_{i,j}}{\sigma_{i,j}},
\end{equation}
where $c_{\rm f}$ is the source covering factor, $S^{\rm c}_{i,j}$ is the unabsorbed continuum flux density, and $\sigma_{i,j}$ is the rms noise. If we consider a Gaussian profile, then the peak optical depth is given in terms of the physical properties of the absorber by 
\begin{equation}\label{equation:optical_depth}
    \tau = 0.034 \left[\frac{N_{\rm HI}}{2 \times 10^{20}\,\mathrm{cm}^{-2}}\right] \left[\frac{T_{\rm s}}{100\,\mathrm{K}}\right]^{-1} \left[\frac{\Delta{v}_{\rm FWHM}}{30\,\mathrm{km}\,\mathrm{s}^{-1}}\right]^{-1}, 
\end{equation}
where $N_{\rm HI}$ is the \mbox{H\,{\sc i}} column density, $T_{\rm s}$ is the spin temperature, and $\Delta{v}_{\rm FWHM}$ the FWHM. We can therefore use  \autoref{equation:signal_to_noise} and \autoref{equation:optical_depth} to calculate the expected peak S/N in a given spectral channel as a function of the physical properties of the absorber, and then use the results of \autoref{figure:completeness} to determine the completeness $C_{i, j}$.

\section{The expected detection yield}\label{section:expected_detection_yield}

To infer the spin temperature, we need to determine how many intervening absorbers we would have expected to detect in the data as a function of $T_{\rm s}$. In the discrete limit, the expected number of detected intervening absorbers is evaluated by the following sum over $J$ sight-lines, each with $I_{j}$ spectral channels,
\begin{multline}\label{equation:expected_detections}
	\mathcal{\overline{\mu}}_{\rm abs}(\theta) = \sum_{j = 1}^{J}\sum_{i=1}^{I_j}\delta{X}_{i,j}\,\int_{N_{\rm min}}^{N_{\rm max}} f(N_{\rm HI}, z_{i,j})\,C_{i,j}(N_{\rm HI}, \theta) \\w_{j}(z_{i,j}+\Delta{z}_{i,j}^{\rm asc})\,\mathrm{d}N_{\rm HI},
\end{multline}	
where $\theta$ is the set of parameters at the absorber that determine detection, $\{T_{\rm s}, c_{\rm f}, \Delta{v}_{\rm FWHM}\}$, $\delta{X}_{i,j}$ is the absorption path length spanned by the $i,j$'th spectral channel, and $f(N_{\rm HI}, z)$ is the column density distribution function, equal to the frequency of absorbers intersecting a given sight-line with a column density between $N_{\rm HI}$ and $N_{\rm HI} + \mathrm{d}N_{\rm HI}$, per unit column density per unit comoving absorption path. The completeness, $C_{i,j}(N_{\rm HI}, \theta)$, is as defined in \autoref{section:completeness}. We integrate over column densities between $N_{\rm min}$ and $N_{\rm max}$, which are selected based on the range of column densities for which we expect the data to be sensitive to 21-cm absorption. The source redshift weighting function, $w_j(z + \Delta{z}_{\rm asc})$, is the probability that the $j$'th source is located at a redshift greater than $z + \Delta{z}_{\rm asc}$. The offset in redshift is used to exclude absorption associated with the host galaxy of the source, and is given by
\begin{equation}
	\Delta{z}_{\rm asc} = (1+z)\Delta{v}_{\rm asc}/c, 
\end{equation}
where $\Delta{v}_{\rm asc} = 3000$\,km\,s$^{-1}$.

To determine $\overline{\mu}_{\rm abs}$ as a function of spin temperature, we marginalise over the velocity width and covering factor
\begin{equation}\label{equation:expected_detections_marginal}
	\overline{\mu}_{\rm abs}(T_{\rm s}) = \iint{\overline{\mu}_{\rm abs}(\theta)\,p(c_\mathrm{f}, \Delta{v}_{\rm FWHM} | T_{\rm s})\,\mathrm{d}(c_\mathrm{f})\,\mathrm{d}(\Delta{v}_{\rm FWHM})},
\end{equation}
where $p(c_\mathrm{f}, \Delta{v}_{\rm FWHM} | T_{\rm s})$ is the joint conditional probability distribution for $c_{\rm f}$ and $\Delta{v}_{\rm FWHM}$. We assume that the covering factor and velocity width are independent, so this can be expressed as a product of the individual conditional probability distributions for these two quantities. 

In the remainder of this section we discuss in further detail the factors and assumptions that are used in \autoref{equation:expected_detections} and \autoref{equation:expected_detections_marginal}.

\subsection{Source redshift}

For the 50 sources in the S20 sample with reliable spectroscopic redshifts, the weighting $w_j(z)$ is simply given by 
\begin{equation}
w_j(z) = \begin{cases}
1 & z < z_{j}^{\rm src} \\
0 & \text{otherwise},
\end{cases}
\end{equation}
where $z_{j}^{\rm src}$ is the redshift of the $j$'th source. Typical uncertainties in the optical spectroscopic redshifts of about 50\,km\,s$^{-1}$ correspond to a fractional uncertainty in each sight-line of less than 0.05\,per\,cent.

For the remaining three sources without spectroscopic redshifts, we use the statistical redshift distribution given by \cite{DeZotti:2010} for bright radio sources ($S_{1.4} > 10$\,mJy) in the Combined EIS-NVSS survey of Radio Sources (CENSORS; \citealt{Brookes:2008}),
\begin{equation}
w_j(z) = \int_{z}^{\infty}{\mathcal{N}_{\rm src}(z^{\prime})\,\mathrm{d}z^{\prime}}\bigg/{{\int_{0}^{\infty}{\mathcal{N}_{\rm src}(z^{\prime})\,\mathrm{d}z^{\prime}}}},
\end{equation}
where 
\begin{equation}
\mathcal{N}_{\rm src}(z) = 1.29 + 32.37\,z - 32.89\,z^{2} \
 + 11.13\,z^{3} - 1.25\,z^{4}.
\end{equation}
Measurement uncertainty in the CENSORS redshift distribution contributes a fractional uncertainty in each sight-line that increases with redshift from approximately 4 per\,cent at $z = 0.37$ to 8\,per\,cent at $z = 1.0$. However, since these sight-lines comprise only 5\,per\,cent of the sample, the uncertainty in the total path length due to the source redshifts is only about 0.3\,per\,cent.

\subsection{Source covering factor}

An absorption line measured from a spectrum is an average over the spatial extent of the unresolved background continuum source. For the interstellar atomic gas in galaxies, distinct structures are evident on linear sizes as small as 100\,pc (\citealt{Braun:2012}), and their column density distribution has been measured in both the nearby Universe, using resolved studies of 21-cm emission (e.g. \citealt{Zwaan:2005a, Braun:2012}), and at cosmological distances using Lyman-$\alpha$ absorption towards compact UV emission from active galactic nuclei (AGNs; e.g. \citealt{Noterdaeme:2012, Neeleman:2016, Rao:2017, Bird:2017}). 

In the case of \mbox{H\,{\sc i}} 21-cm absorption the background radio source can extend well beyond the AGN, so that at the intervening galaxy a large fraction of the flux density is distributed on scales larger than that probed by Lyman-$\alpha$ absorption. The effect of averaging over extended radio sources is to distribute the 21-cm absorbers to lower optical depths than would be expected given the distribution measured at higher spatial resolution (\citealt{Braun:2012}). The impact on our model is to over predict how many 21-cm absorbers we expect to detect. It is therefore useful to define a source covering factor, $c_{\rm f}$, that corrects for the unknown areal fraction of the source flux density that is subtended by a foreground absorbing structure of putative constant optical depth (see \autoref{equation:signal_to_noise}). 

Ideally we need information about the mas-structure of each source at sub-GHz frequencies, in order to fully understand the expected effect of source size for 21-cm absorption (e.g. \citealt{Braun:2012}; K14). In the absence of this information for our sample, we instead extrapolate from what we understand about the source morphology at other wavelengths. S20 selected sources from the AT20G catalogue (\citealt{Murphy:2010}) that have flux densities greater than 500\,mJy at 20\,GHz and 1.5\,Jy at 1.4\,GHz/843\,MHz, thereby including a high fraction of radio-loud QSOs with compact radio emission. Their analysis of very long baseline interferometric (VLBI) imaging at 5 -- 15\,GHz in the literature indicated that these sources typically have 20 -- 60\,per\,cent of their flux density in components smaller than $\sim 10$\,mas, corresponding to projected sizes between 50 and 80\,pc for the redshift range used in this survey. They note that this is consistent with the results of \cite{Horiuchi:2004}, who found that for a larger complete and flux-density limited sample of 303 radio sources at 5\,GHz, about 50\,per\,cent of the flux density is typically contained in a 10\,mas component, with 20\,per\,cent from a radio core of average size 0.2\,mas. 

If we assume that the ratio of the 10-mas flux density to the total flux density is a reasonable proxy for the covering factor, then based on the above results we might expect that a typical value is $\langle{c_{\rm f}}\rangle \approx 0.5$. We note that this also assumes that the source structure at 5\,GHz is representative of the structure at sub-GHz frequencies. However, \cite{Kanekar:2009} (see also K14) carried out a VLBI imaging study at sub-GHz frequencies of the radio-loud quasars in their DLA sample, using the ratio of core-to-total flux density as a proxy for the covering factor. \cite{Allison:2016} used a two-tailed Kolmogorov-Smirnov (KS) test to show that the hypothesis that covering factors obtained by Kanekar et al. are drawn from a $0 - 1$ uniform distribution is true at the level of $p = 0.05$, but not for $p = 0.01$. Visual inspection of the distribution shows that there may be a paucity of quasars in the Kanekar et al. sample for covering factors less than $c_{\rm f} \sim 0.2$\, which is consistent with the \cite{Horiuchi:2004} result that 20\,per\,cent of the flux density in radio-loud AGN selected at 5\,GHz is within the sub-mas radio core. 

Based on these results, we marginalise the expected number of absorbers over the covering factor by drawing randomly from a uniform prior between $c_{\rm f} = 0$ and 1. We further assume that the covering factor is not conditional on the spin temperature. It is possible that the true distribution of covering factors may be skewed to lower values than uniform, in which case we would underestimate the expected detection yield and therefore overestimate the inferred spin temperature. 

\subsection{Velocity width}

\begin{figure}
\begin{center}
\includegraphics[width=\columnwidth]{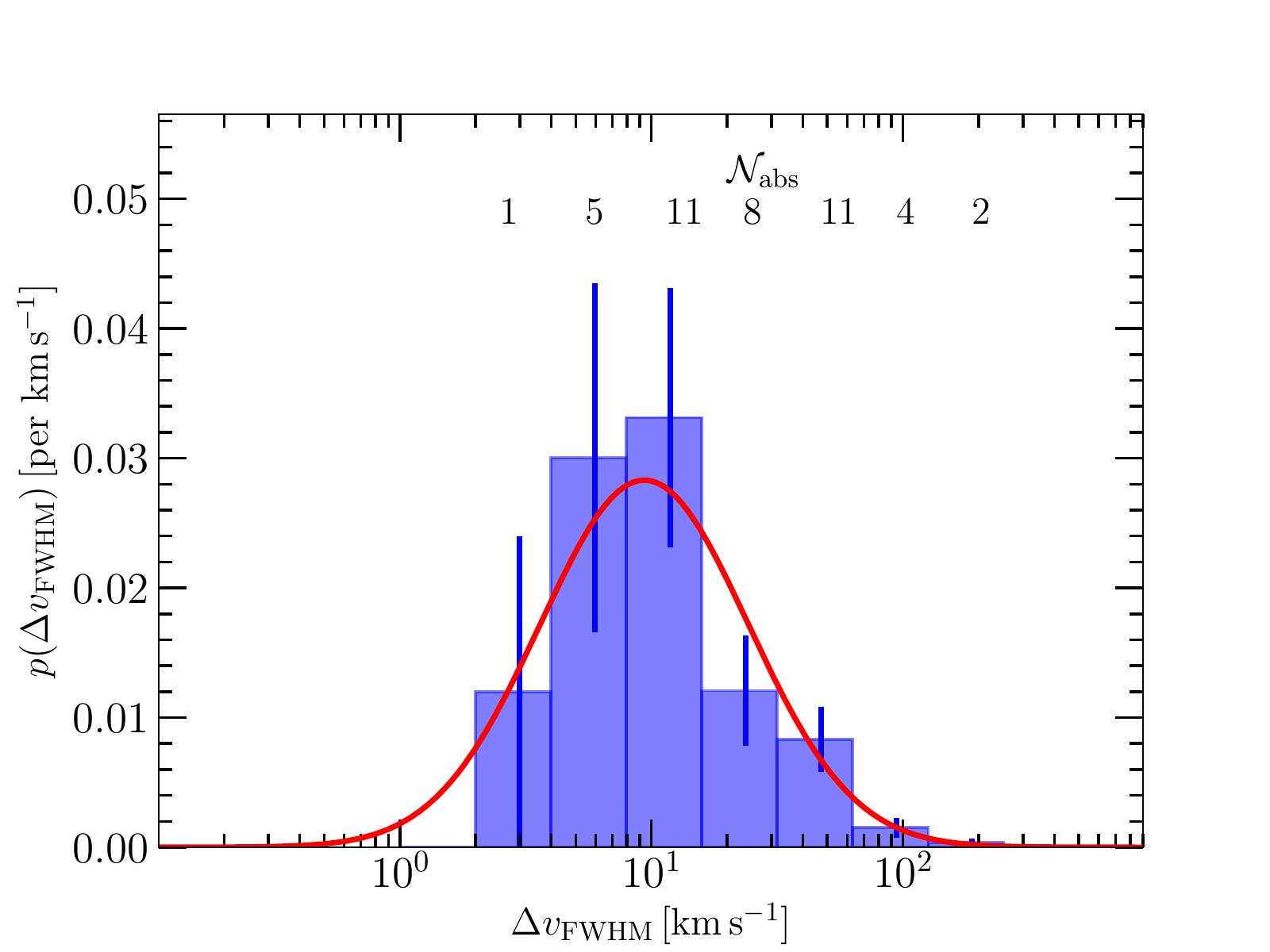}
\caption{The sample distribution of velocity widths in the literature, for intervening 21-cm absorbers at $z > 0.1$ (blue histogram). Errorbars denote the standard deviation given by $\sqrt{\mathcal{N}_{\rm abs}}$. The red line is a log-normal fit to the data, from which we draw our prior distribution of widths.}\label{figure:width_dist}
\end{center}
\end{figure}

Sensitivity to a resolved 21-cm absorption line of fixed equivalent width is inversely related to the square-root of its width. We marginalise the expected number of detections over the distribution of velocity widths for a sample of detected intervening 21-cm absorbers in the literature\footnote{References for the literature sample of velocity widths shown in \autoref{figure:width_dist}: \citet*{Briggs:2001}; \citet*{Carilli:1993}; \citet*{Chengalur:1999}; \citet{Chengalur:2000, Curran:2007, Davis:1978, Ellison:2012, Gupta:2009, Gupta:2012, Gupta:2013}; \citet{Kanekar:2001b,Kanekar:2003b};  \citet{Kanekar:2001c, Kanekar:2006, Kanekar:2009a, Kanekar:2013, Kanekar:2014a};  \citet{Kanekar:2003a},  \citet*{Kanekar:2007};  \citet*{Kanekar:2014b};  \citet{Lane:2001, Lovell:1996, York:2007, Zwaan:2015}.}. We assume that this sample distribution is drawn from a sufficiently large range of observations with different sensitivities and spectral resolutions to be representative of the population and not censored by the underlying sensitivity to velocity width. This is shown in \autoref{figure:width_dist}, along with a log-normal fit to the data, from which we draw random line widths. 

By using this single distribution for the velocity width, we have assumed that it is not conditional on the spin temperature. This is true for absorption lines where the dominant broadening mechanism is from bulk rotational or turbulent motion of the gas, but not when thermal broadening is important. Since the spin temperature is either equal to or less than the gas kinetic temperature, depending on the dominant mechanism for excitation of the 21-cm line (e.g. \citealt{Purcell:1956, Field:1958, Field:1959, Bahcall:1969, Liszt:2001}), this places a lower limit constraint on the velocity width. We therefore apply the following additional constraint on the allowed range of velocity widths
\begin{equation}
	\Delta{v}_{\rm FWHM} \geq \sqrt{8\,\ln(2)\,\frac{k_{\rm B}}{m_{\rm H}}\,T_{\rm s}},
\end{equation}
where $k_{\rm B}$ is the Boltzmann constant and $m_{\rm H}$ is the mass of a hydrogen atom. Note that for typical \mbox{H\,{\sc i}} spin temperatures in the range 100 -- 1000\,K, this corresponds to lower bounds on the FWHM between 2 and 7\,km\,s$^{-1}$, which is consistent with the literature distribution.

\subsection{Column density distribution function}

The column density distribution function, $f(N_{\rm HI})$, gives the frequency of absorbers intercepting a given sight-line with a column density between $N_{\rm HI}$ and $N_{\rm HI} + \mathrm{d}N_{\rm HI}$, per unit column density per unit comoving absorption path. In \autoref{equation:expected_detections} we integrate over column densities that span equivalent DLA systems ($N_{\rm min} = 2 \times 10^{20}$\,cm\,$^{-2}$, $N_{\rm max} = \infty$), which are thought to have sufficient warm neutral gas to temperature-shield and produce CNM (\citealt{Kanekar:2011}). The four intervening absorbers detected in this sample all have integrated optical depths that correspond to column densities in the equivalent DLA range for $T_{\rm s} > 100$\,K and $c_{\rm f} < 1$ (S20), suggesting that this is the range of column densities for which our data sensitive. The expected number of detected 21-cm absorbers should therefore be considered as those with column densities in the equivalent DLA range. 

Precise measurements of $f(N_{\rm HI})$ have been obtained using 21-cm emission line surveys in the local Universe (e.g. \citealt{Zwaan:2005a, Braun:2012}) and DLAs at cosmological distances (e.g. \citealt{Noterdaeme:2012, Neeleman:2016, Rao:2017, Bird:2017}). Searches for DLAs at UV-wavelengths using archival \textit{HST} data by \cite{Neeleman:2016} and \cite{Rao:2017} are most closely matched in redshift to our data, but the sample sizes are insufficient to provide precise measurements of $f(N_{\rm HI})$ across the full range of equivalent DLA column densities. We therefore linearly interpolate between the results of \cite{Zwaan:2005a} and \cite{Braun:2012} at $z = 0$, and \cite{Bird:2017} at $z = 2$, who obtained the most precise DLA measurement yet using a catalogue generated by a Gaussian Process (GP) method from the  Sloan Digital Sky Survey III Data Release 12 (\citealt{Garnett:2017}). 

\subsubsection{The low-$z$ $f(N_{\rm HI})$: spatial resolution and self-absorption}

In the low-$z$ local Universe, the column density distribution functions measured by \cite{Zwaan:2005a} and \cite{Braun:2012} disagree significantly, the former being steeper than expected at high column densities for a random oriented gas disc. The measurement by \cite{Zwaan:2005a} was carried out using HI emission images from a sample of 355 galaxies observed in the Westerbork HI Survey of Irregular and Spiral Galaxies (WHISP; \citealt{vanderHulst:2001}) with a maximum linear resolution of about 1.3\,kpc. In contrast, \cite{Braun:2012} only looked at the \mbox{H\,{\sc i}} images of three galaxies in the Local Group (M31, M33 and the Large Magellanic Cloud), but in much greater detail at 100\,pc resolution and with thousands of independent sight lines.

There are several possible reasons for the differences seen in these two measurements. First, the lower resolution of the study by \cite{Zwaan:2005a} will re-distribute small-scale structures to lower column densities, and thus steepen $f(N_{\rm HI})$. \cite{Braun:2012} suggested that this is broadly consistent with their data once smoothed to a similar resolution, but not reproduced in detail. Secondly, by modelling each emission line profile in detail, \cite{Braun:2012} showed that 21-cm self-absorption can have a significant effect in reducing high column densities in the range $22 < \log_{10}(N_{\rm HI}) < 23$. However, we note that despite the exquisite spatial detail and self-consistency within the Local Group sample, it is not yet clear how representative it is of the \mbox{H\,{\sc i}} distribution in the larger galaxy population at $z = 0$. 

Given the discrepancy between these measurements, we calculate the expected number of 21-cm absorbers using both distributions and use the resulting difference as the standard error (with a median of 27\,per\,cent).

\subsubsection{Dust obscuration and redenning in DLA sight-lines}

The possibility that a significant fraction of quasars with dusty intervening absorbers are missing from flux-limited optical samples has long been the subject of vigorous investigation. If true, then these optical surveys would underestimate the number density of high-$N_{\rm HI}$ DLAs at $z > 2$ and there would be a corresponding steepening of the column density distribution function that we use here.

 Until recently there was no strong evidence of a large missing fraction of optical DLAs; in a combined Bayesian analysis of the optical and radio constraints \cite{Pontzen:2009} found that only 7\,per\,cent of DLAs are expected to be missing from optically-selected samples due to dust obscuration, with at most 19\,per\,cent missing at 95\,per\,cent confidence. However, \cite{Krogager:2019} have recently considered the colour selection used in SDSS-II Data Release 7, in addition to the magnitude limit, and found that the fraction of missing DLAs could be as much as 28\,per\,cent at $z \approx 3$  and 42\,per\,cent at $z \approx 2$, for the lowest dust-to-metal ratio of $\log_{10}\kappa_{\rm Z} = -21.1$. 
 
 The column density distribution function we use here was measured by \cite{Bird:2017} using spectra from SDSS-III DR12, for which a similar analysis has yet to be carried out. \cite{Krogager:2019} state that although the selection criteria used for SDSS-III is more complex, it may still be susceptible to the same biases in SDSS-II. In the absence of further information, we calculate the expected number of 21-cm absorbers in our data assuming that the column density distribution function of \cite{Bird:2017} may be systematically low by 20\,per\,cent, using the resulting difference as the standard error (with a median of 9.7\,per\,cent).
   
 \subsection{Absorption path length}

\begin{figure}
\centering
\includegraphics[width=0.45\textwidth]{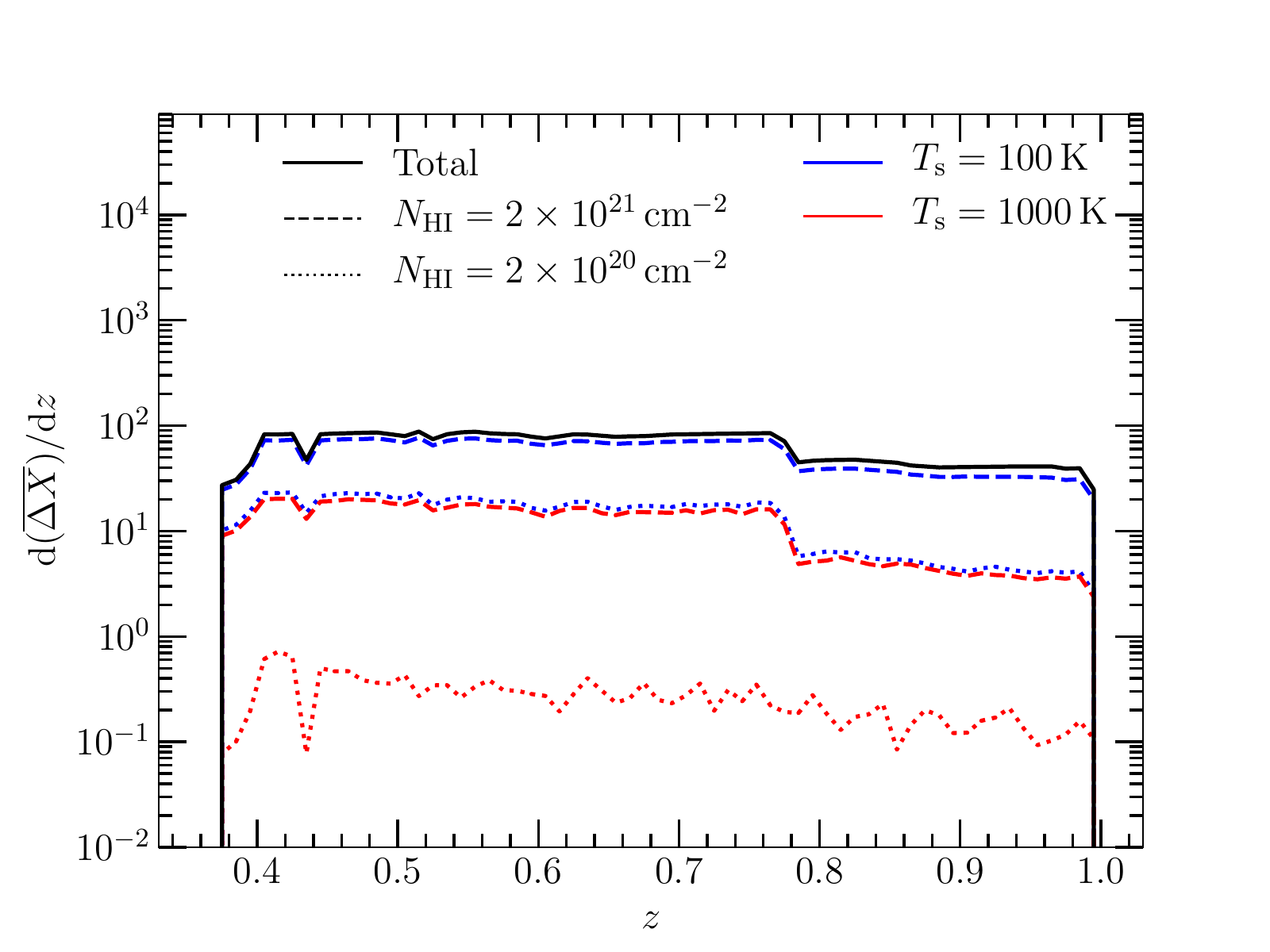}
\caption{The change in absorption path length with redshift. Lines are shown for the total interval spanned by the survey (regardless of spin temperature), and for intervals sensitive to DLAs ($2 \times 10^{20}$\,cm$^{-2}$) and super-DLAs ($2 \times 10^{21}$\,cm$^{-2}$), for spin temperatures $T_{\rm s} = 100$ and 1000\,K.}
\label{figure:deltax_vs_z}
\end{figure} 

The absorption path length defines the comoving interval of the survey that is sensitive to intervening absorbers. The infinitesimal path length element $\mathrm{d}{X}$ spanned by a redshift element $\mathrm{d}{z}$ is given analytically by 
\begin{equation}
 \mathrm{d}{X} = \mathrm{d}{z}\,(1+z)^{2}\,E(z)^{-1},
\end{equation}
where
\begin{equation}
\mathrm{d}{z} = (1+z)\frac{\mathrm{d}\nu}{\nu},
\end{equation}
\begin{equation}
z = \frac{\nu_{\rm HI}}{\nu} - 1,
\end{equation}
\begin{equation}
 E(z) = {\sqrt{(1+z)^{3}\,\Omega_{\rm m} + (1+z)^{2}\,(1 - \Omega_{\rm m} - \Omega_{\Lambda}) + \Omega_{\Lambda}},}
\end{equation} 
$\nu$ is the corrected observed frequency, and $\nu_{\rm HI}$ is the rest frequency of the \mbox{H\,{\sc i}} 21-cm line, equal to 1420.40575177\,MHz (\citealt{Hellwig:1970}). 

In the discrete limit, We can estimate the expected total absorption path length sensitive to a given column density by evaluating the following sum over $J$ sight lines, each with $I_j$ channels
\begin{equation}\label{equation:total_absorption_path}
    \overline{\Delta{X}}(N_{\rm HI}) = \sum_{j=1}^{J}\sum_{i=1}^{I_{j}} {C_{i,j}(N_{\rm HI}, \theta)\,w_j(z_{i,j} + \Delta{z}_{i,j}^{\rm asc})\,\delta{X}_{i,j}},
\end{equation}
where the completeness and source redshift weighting are as defined above. Again marginalising over the above distributions for $\Delta{v}_{\rm FWHM}$ and $c_{\rm f}$, and assuming a fiducial $T_{\rm s} = 100$\,K, the expected total path length sensitive to DLA column densities ($N_{\rm HI} > 2 \times 10^{20}$\,cm$^{-2}$) is $\overline{\Delta{X}} = 8.7$ $(\overline{\Delta{z}} = 4.7)$ and for super-DLAs ($N_{\rm HI} > 2 \times 10^{21}$\,cm$^{-2}$) is $\overline{\Delta{X}} = 35$  $(\overline{\Delta{z}} = 19)$. For a higher spin temperature of $T_{\rm s} = 1000$\,K, these intervals decrease to $\overline{\Delta{X}} = 0.17$ $(\overline{\Delta{z}} = 0.092)$ and $\overline{\Delta{X}} = 7.5$ $(\overline{\Delta{z}} = 4.1)$, respectively. In \autoref{figure:deltax_vs_z} we show the differential absorption path length, which gives the sensitivity to 21-cm absorbers as a function of redshift. Since observations with the more sensitive ASKAP-12 telescope were undertaken at higher frequencies, this sensitivity function is skewed to lower redshifts.  

\section{Results}

\begin{figure}
\begin{center}
\includegraphics[width=\columnwidth]{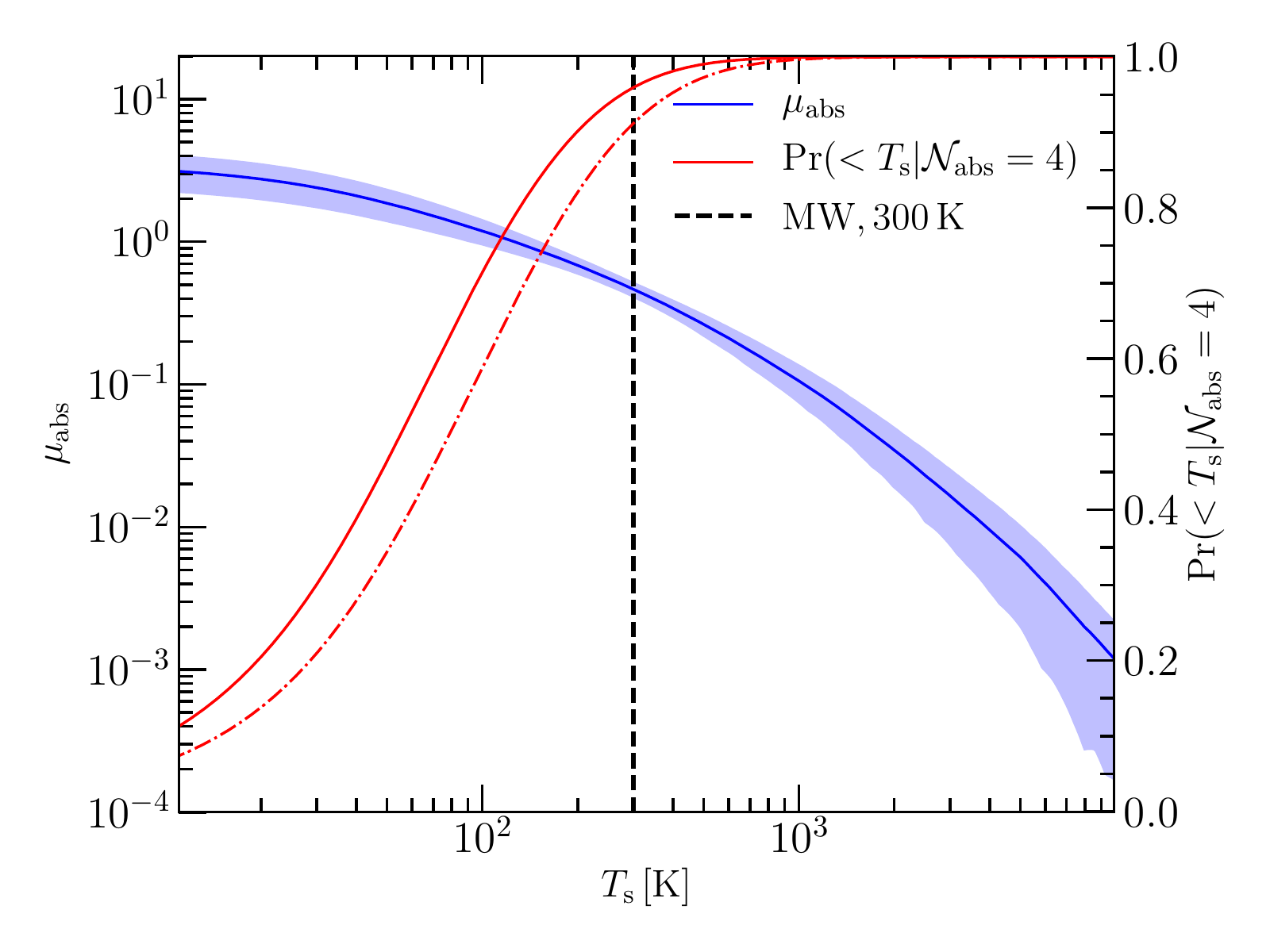}
\caption{The expected number of detected intervening 21-cm absorbers (blue line) and 68\,per\,cent uncertainty interval (shaded region), and the corresponding cumulative posterior probability for the spin temperature, conditional on detecting 4 absorbers (red line). The two red lines correspond to different choices of prior: $p(\mu_{\rm abs}) \propto 1/\sqrt{\mu_{\rm abs}}$ (solid) and $1/\mu_{\rm abs}$ (dot-dashed).} \label{figure:nabs_vs_tspin}
\end{center}
\end{figure}

\subsection{A statistical measurement of the spin temperature}

We show the expected number of detected intervening 21-cm absorbers as a function of spin temperature in \autoref{figure:nabs_vs_tspin}. The 68\,per\,cent uncertainty is also given, equal to the quadrature sum of the standard error due to measurement uncertainty (assumed in total to be about 10\,per\,cent), systematic differences between the nearby 21-cm emission line surveys, and the dust obscuration in DLA surveys. The median uncertainty in $\mu_{\rm abs}$ is about 30\,per\,cent, but increases significantly at high spin temperatures due to uncertainty in the number of high column density systems that dominate at these sensitivities. The expected number of detections at the mass-weighted harmonic mean spin temperature for the Milky Way ISM ($\approx 300$\,K; \citealt{Murray:2018}) is $0.46 \pm 0.06$, which is notably less than the four detections obtained by S20 and indicates possible tension with the data. 

We determine the posterior probability density function\footnote{In the notation used here, $p(x)$ is the probability density function for a continuous variable $x$, $\mathrm{Pr}(< x)$ is the probability mass that a continuous variable has a value less than $x$, and $\mathrm{Pr}(y)$ is the probability mass function for a discrete variable y.} for the spin temperature using Bayes' Theorem,
\begin{equation}\label{equation:tspin_probability_density}
	p(T_{\rm s} | \mathcal{N}_{\rm abs}) = \frac{\mathrm{Pr}(\mathcal{N}_{\rm abs} | T_{\rm s})\,p(T_{\rm s})}{\mathrm{Pr}(\mathcal{N}_{\rm abs})},
\end{equation}
where $\mathrm{Pr}(\mathcal{N}_{\rm abs} | T_{\rm s})$ is the likelihood function, $p(T_{\rm s})$ is the prior probability density, and $\mathrm{Pr}(\mathcal{N}_{\rm abs})$ is the normalising constant known as the marginal likelihood or model evidence. In \autoref{section:expected_detection_yield} we obtained the probability density for the detection yield ($\mu_{\rm abs}$) conditional on the spin temperature, $p(\mu_{\rm abs} | T_{\rm s})$. We model this as a normal distribution with mean and standard deviation as shown in \autoref{figure:nabs_vs_tspin}. These results are then used to determine the probability of detecting $\mathcal{N}_{\rm abs}$ as a function of spin temperature by evaluating the following integral,
\begin{equation}
	\mathrm{Pr}(\mathcal{N}_{\rm abs} | T_{\rm s}) = \int\,\mathrm{Pr}(\mathcal{N}_{\rm abs} | \mu_{\rm abs})\,p(\mu_{\rm abs} | T_{\rm s})\,\mathrm{d}\mu_{\rm abs},
\end{equation}
where the likelihood function for $\mu_{\rm abs}$ is modelled as a Poisson distribution given by
\begin{equation}
	\mathrm{Pr}(\mathcal{N}_{\rm abs} | \mu_{\rm abs}) = \frac{{\mu_{\rm abs}}^{\mathcal{N}_{\rm abs}}	}{\mathcal{N}_{\rm abs}!}\mathrm{e}^{-\mu_{\rm abs}}.
\end{equation}

To determine the posterior probability density given in \autoref{equation:tspin_probability_density} we also need to choose an appropriate prior, $p(T_{\rm s})$. For smaller surveys, where the number of detections are fewer and hence the likelihood is less constraining, this choice can have a significant affect on the inferred spin temperature. Since we have no strong prior information on the value of the spin temperature, we choose a non-informative prior. In the context of the Poisson process being considered here, the spin temperature prior is given in terms of the detection yield by
\begin{equation}
	p(T_{\rm s}) = \int{p(T_{\rm s} | \mu_{\rm abs})\,p(\mu_{\rm abs})\,\mathrm{d}\mu_{\rm abs}}, 
\end{equation}
where $p(T_{\rm s} | \mu_{\rm abs})$ is just the inverse of the above-mentioned normal distribution, and $p(\mu_{\rm abs})$ is the prior for the detection yield. We choose the non-informative Jeffreys prior, $p(\mu_{\rm abs}) \propto 1/\sqrt{\mu_{\rm abs}}$, which corresponds to invariance of the detection yield under a change of parameterisation (\citealt{Jeffreys:1946}).  This is an improper prior, but valid for a finite range of spin temperatures. We choose $0.1\,\mathrm{K} < T_{\rm s} < 10^{5}\,\mathrm{K}$, covering all reasonable values of the spin temperature expected from our understanding of the physical conditions in the interstellar medium of galaxies (e.g. \citealt{Wolfire:2003}). An equally suitable choice of non-informative prior would be $p(\mu_{\rm abs}) \propto 1/\mu_{\rm abs}$ (e.g. \citealt{Jeffreys:1961, Novick:1965, Villegas:1977}), which we also consider here in our results. 
 
We can now estimate the posterior probability for the spin temperature given that four intervening 21-cm absorbers were detected in the S20 data. In \autoref{figure:nabs_vs_tspin} we plot the resulting cumulative probability function, $\mathrm{Pr}(< T_{\rm s} | \mathcal{N}_{\rm abs})$ , calculated by integrating the posterior probability density given by \autoref{equation:tspin_probability_density}. We find that $T_{\rm s} < 274 $\,K with 95\,per\,cent probability, increasing to 387\,K for the above alternative choice of non-informative prior. 

\subsection{Interpreting the measurement}

Since the 21-cm line optical depth is inversely proportional to the spin temperature, the value inferred from such a measurement is a column-density-weighted harmonic mean over the line-of-sight \mbox{H\,{\sc i}} present in each thermally distinct phase, 
\begin{equation}
	T_{\rm s} = N_{\rm HI}\,\left[\sum_{i}\frac{N_{i}}{T_{i}}\right]^{-1},
\end{equation}
where $N_{\rm HI} = \sum_{i}{N}_{i}$.  In the case of a simple two-phase model, where the spin temperatures of the individual phases are known, then the measurement of  $T_{\rm s}$ can be inverted to infer their relative fraction. In practice the coldest phase dominates the measurement and this expression reduces to a simple ratio of the cold fraction and temperature. However, in general any predictive model of the multi-phased ISM can be tested by measuring $T_{\rm s}$.

In the Milky Way ISM there are three observed thermally distinct phases: the cold (CNM; $T_{\rm CNM} \sim 100$\,K), unstable (UNM; $T_{\rm UNM} \sim 500$\,K) and warm neutral medium (WNM; $T_{\rm WNM} \sim 10\,000$\,K), with total mass fractions of approximately 28, 20 and 52\,per\,cent respectively (\citealt{Heiles:2003, Murray:2018}). The mass-weighted harmonic mean spin temperature over the Milky Way is therefore approximately 300\,K, and is seen to be constant with galactocentric radius outside of the solar circle (\citealt{Dickey:2009}). This suggests that the phases are sufficiently mixed that random DLA-like sight-lines observed through our Galaxy would give on average a measurement of $T_{\rm s} \sim 300$\,K.  

 Given that the relative fraction of neutral phases depends on the gas-phase metallicity, dust abundance and ambient UV field of individual galaxies (\citealt{Wolfire:2003}), we expect $T_{\rm s}$ to vary within the population. Such variation is seen within the Local Group, where the Small Magellanic Cloud has a much smaller fraction of CNM than other members, although this is somewhat mitigated by a correspondingly lower CNM temperature (\citealt{Dickey:2000}). Likewise we expect to see variation in the $T_{\rm s}$ inferred from DLAs simply due to variance created by sampled sight-lines; indeed a high intrinsic scatter is seen in the spin temperature of individual DLAs within a given redshift range (see \autoref{figure:tspin_vs_z} and K14). 
 
 Despite this variance in the $T_{\rm s}$ for individual objects, we can test for systematic changes in the physical conditions of \mbox{H\,{\sc i}} gas with redshift. For the individual DLAs in the sample of K14, an estimate of the population $T_{\rm s}$ can be obtained by taking the sample $N_{\rm HI}$-weighted harmonic mean. Likewise, the value of $T_{\rm s}$ we infer from \autoref{equation:tspin_probability_density} is the $N_{\rm HI}$-weighted harmonic mean over the DLA population within the redshift range spanned by the survey, and can therefore be usefully compared with direct measurements at other redshifts.
 
\subsection{Including results from the FLASH GAMA23 survey}

\begin{figure}
\centering
\includegraphics[width=0.45\textwidth]{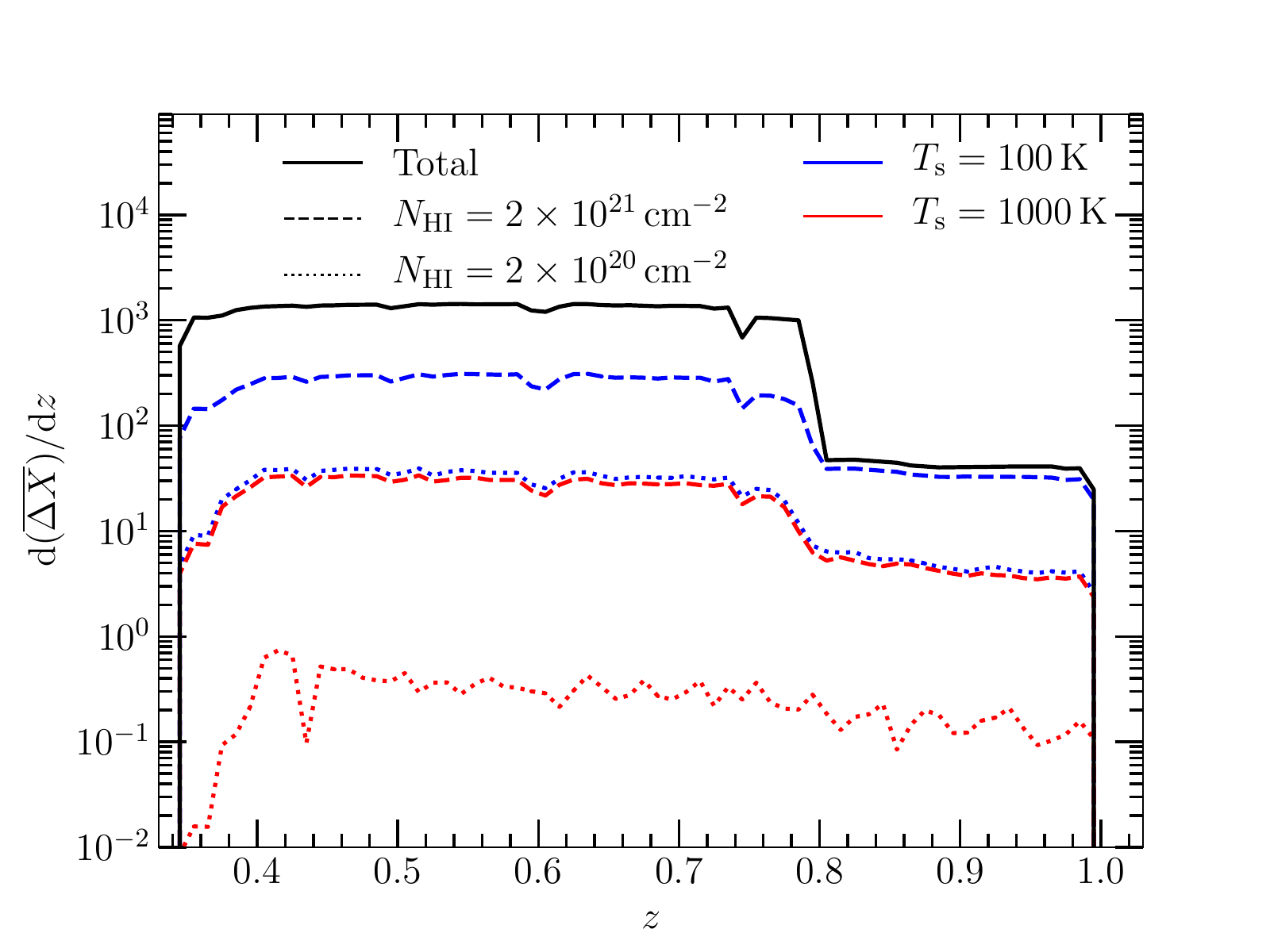}
\caption{As \autoref{figure:deltax_vs_z}, but including data from the wide field survey of the GAMA23 field, undertaken as a FLASH early science projct (A20).}
\label{figure:deltax_vs_z_withGAMA23}
\end{figure} 
 
As previously mentioned in \autoref{section:reliability}, A20 (\citealt{Allison:2020}) recently reported results from a wide-field \mbox{H\,{\sc i}} 21-cm absorption survey of the GAMA\,23 field (\citealt{Liske:2015}), undertaken as a FLASH early science project with ASKAP-12. They searched for \mbox{H\,{\sc i}} absorption towards 1253 radio sources at a median rms noise level of 3.2\,mJy\,beam$^{-1}$ per 18.5kHz, covering redshifts between $z = 0.34$ and 0.79 over a sky area of approximately 50\,$\deg^{2}$. In a purely blind search of the data Allison et al. did not detect any absorption lines, but by cross-matching radio sight lines with known optically-selected galaxies they did find an absorber associated with gas in the outer regions of an early type galaxy. Allison et al. calculated the expected detection rate as a function of the spin temperature to covering factor, but unlike the method described here, assumed a constant line width $\Delta{v}_{\rm FWHM} = 30$\,km\,s$^{-1}$ and a fixed completeness threshold corresponding to a S/N of 5.5 in a single channel. They reported that if the typical spin temperature to covering factor ratio at these redshifts is equal to 300\,K, then the probability of detecting no absorbers is about 64\,per\,cent. 

It is straight forward to update our statistical measurement of the spin temperature to include this additional information, given that the linear combination of independent Poisson processes is itself a Poisson process with mean equal to the sum of individual means. We re-analyse the data using the method described in this paper, and the completeness simulations carried out by A20 to a reliability threshold of $\ln{B} > 16$. The total absorption path length in the combined S20 and A20 surveys can be estimated using \autoref{equation:total_absorption_path}, which for $T_{\rm s} = 100$\,K we find that $\overline{\Delta{X}} = 15$ ($\overline{\Delta{z}} = 8.2$) for DLA column densities, and $\overline{\Delta{X}} = 120$ ($\overline{\Delta{z}} = 69$) for super-DLA column densities. Likewise, for $T_{\rm s} = 1000$\,K the intervals are $\overline{\Delta{X}} = 0.18$ ($\overline{\Delta{z}} = 0.097$) and $\overline{\Delta{X}} = 13$ ($\overline{\Delta{z}} = 7.0$), sensitive to DLAs and super-DLAs, respectively. In \autoref{figure:deltax_vs_z_withGAMA23} we show the differential absorption path length as a function of redshift, giving the sensitivity function to 21-cm absorbers over the interval spanned by the combined surveys. Given that the A20 survey spanned redshifts between $z = 0.34$ and $0.79$,  there is a corresponding increase in the path length over this range.

\begin{figure}
\begin{center}
\includegraphics[width=\columnwidth]{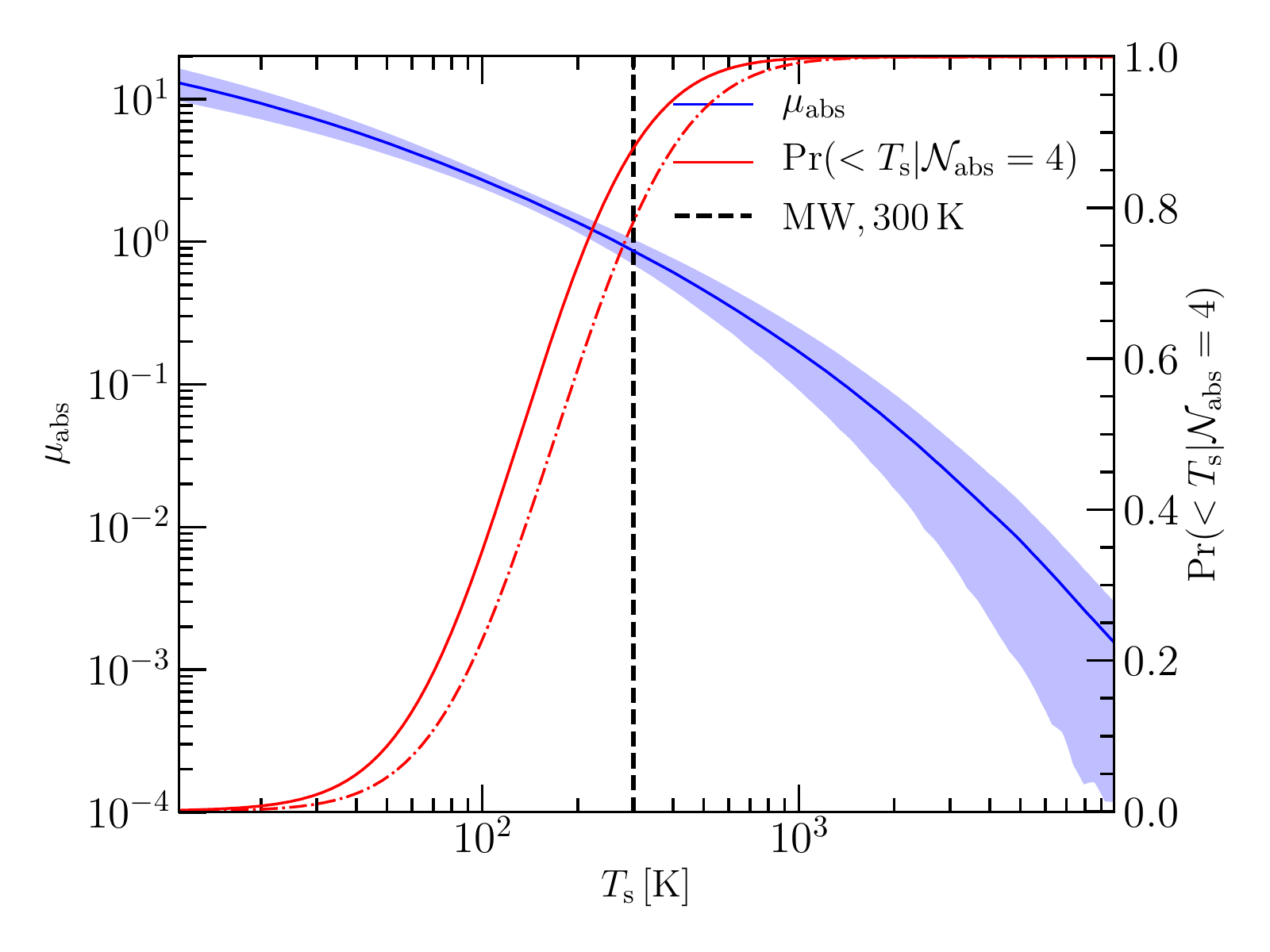}
\caption{As \autoref{figure:nabs_vs_tspin}, but including data from the wide field survey of the GAMA23 field, undertaken as a FLASH early science project (A20).} \label{figure:nabs_vs_tspin_withGAMA23}
\end{center}
\end{figure}

In \autoref{figure:nabs_vs_tspin_withGAMA23} we show the combined number of 21-cm absorbers expected to be detected in both surveys as a function of spin temperature, and the updated spin temperature probability conditional on detecting 4 absorbers. For the Jeffreys prior we find that $T_{\rm s} < 420$\,K with 95\,per\,cent probability, increasing to 567\,K for the alternative prior. Given that the blind survey of the GAMA\,23 field did not yield any further detections of intervening 21-cm absorbers, our upper limit on the harmonic mean spin temperature does increase, as one would expect. However, we caution that the sources in the GAMA\,23 survey were selected purely based on their total flux density, and so the distribution of covering factors for those sources could be significantly skewed to lower values than the uniform prior assumed here. It is therefore likely that the spin temperature is lower than this limit and closer to that determined from just the survey of S20.

\section{Discussion}

\subsection{Comparison with other 21-cm absorption surveys} 

\subsubsection{Direct measurements of $T_{\rm s}$ in individual DLAs}

There have been several 21-cm absorption line surveys of known individual DLAs with the goal of directly measuring the spin temperature, although it should be noted that these too suffer from uncertainty in the source covering factor. These targeted surveys have been compiled and summarised into a single study by K14, who found that DLAs at $z > 2.4$ have a statistically (4\,$\sigma$ significance) different  distribution of spin temperatures to those at lower redshifts, and are on average higher. They also found evidence (3.5\,$\sigma$ significance) of an anti-correlation between the spin temperature and metallicity in DLAs. These results are expected if the DLAs observed in the earlier Universe are on average more relatively metal poor (e.g. \citealt{Rafelski:2012, Cooke:2015, DeCia:2018}) and therefore lacked sufficient coolants in the gas to form a significant fraction of CNM via fine structure cooling. 

\begin{figure}
\begin{center}
\includegraphics[width=\columnwidth]{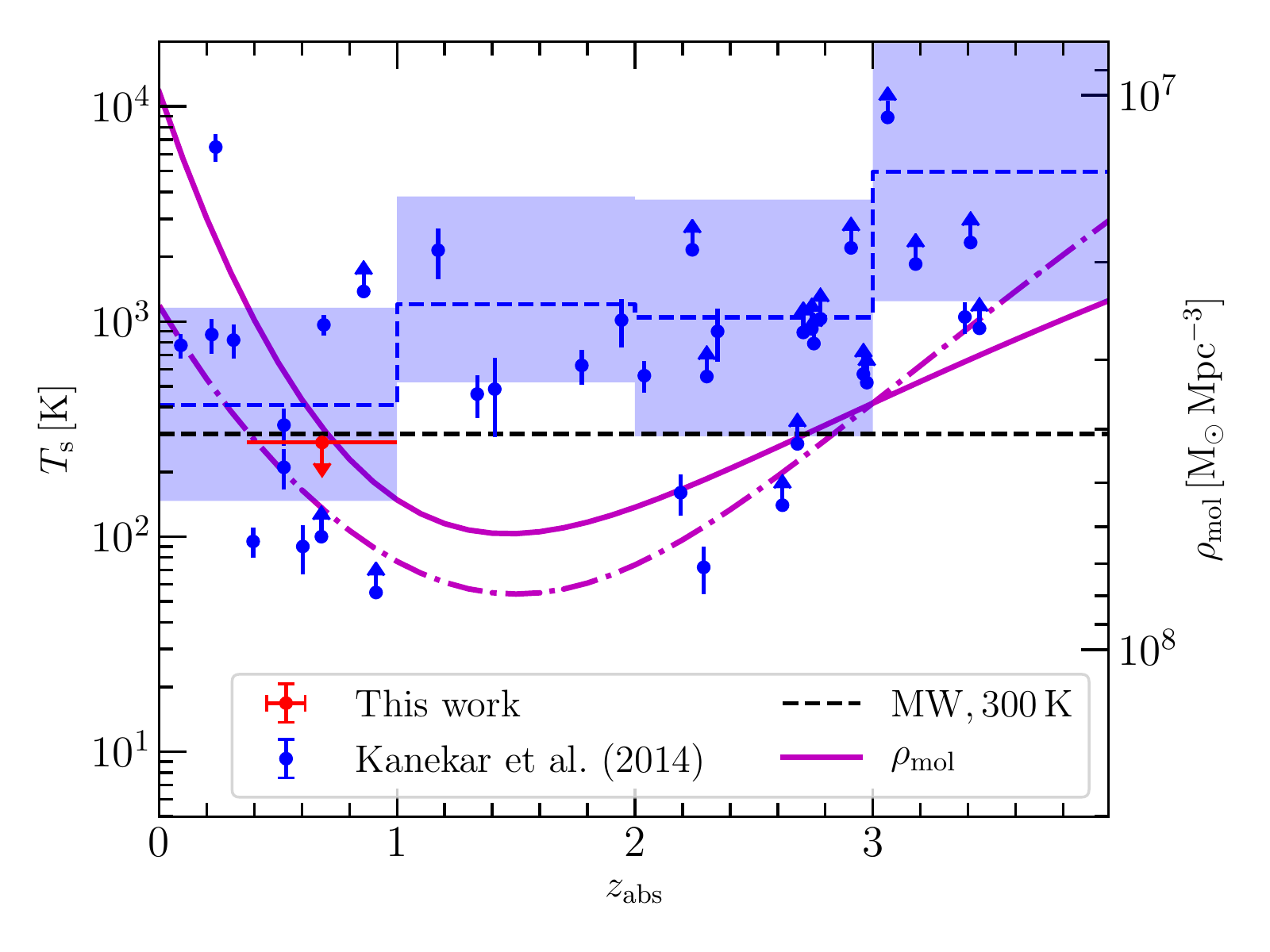}
\caption{The $N_{\rm HI}$-weighted harmonic mean spin temperature versus absorber redshift for this work (red point;  95\,per\,cent upper limit) and individual DLAs in the literature (blue points; K14). The blue dashed line (median) and shaded region (95\,per\,cent confidence interval) are an estimate of the $N_{\rm HI}$-weighted harmonic mean in $\Delta{z} = 1$ bins from the K14 sample (using survival analysis to include limits). Also shown are the evolution of the molecular gas density predicted by the best fitting curve to observational data by \citet{Walter:2020} (solid magenta curve) and the depletion-time model of \citet{Tacconi:2020} (dot-dashed magenta curve).}\label{figure:tspin_vs_z}
\end{center}
\end{figure}

In \autoref{figure:tspin_vs_z} we plot as a function of redshift the 95\,per\,cent upper limit on the $N_{\rm HI}$-weighted harmonic mean spin temperature from our statistical measurement, and the direct measurements from the DLA sample of K14. Using the Kaplan-Meier estimator of the survival function to include lower limits\footnote{We make use of the \textsc{LifeLines} survival analysis package (\url{https://github.com/CamDavidsonPilon/lifelines}).}, we calculate the median and 95\,per\,cent confidence interval for the $N_{\rm HI}$-weighted harmonic mean of the K14 sample in redshift bins of $\Delta{z} = 1$. The results show an evolution with redshift towards larger spin temperatures that is consistent with the conclusions of K14, and in the lowest redshift bin is consistent with our upper limit. 

 \subsubsection{GBT survey of compact sources}
 
In a recent Green Bank Telescope (GBT) survey for intervening 21-cm absorbers towards 252 radio sources, spanning redshifts between $z = 0$ and $2.74$, \cite{Grasha:2020} reported ten detections in a total comoving absorption path length sensitive to DLAs of $\Delta{X} \sim 155$. By comparing their measurement of the cosmological \mbox{H\,{\sc i}} mass density from 21-cm absorbers with prior measurements at the same redshifts, they obtained a mean spin temperature to covering factor ratio of $T_{\rm s}/c_{\rm f} \sim 175$\,K. For all values of $c_{\rm f} \leq 1$ their estimate of the mean spin temperature is consistent with our 95\,per\,cent upper limit on the harmonic mean. However, since Grasha et al. did not provide a confidence interval for this measurement we cannot yet draw a robust statistical comparison. We note that the total absorption path length covered by the GBT survey is a factor $\sim 4$ greater than our data (and is more sensitive to lower column densities), but achieved only  a factor $\sim 2$ greater detection yield. This apparent inconsistency in outcomes between the surveys, and the inferred spin temperature, can be resolved by applying our method to these data in future work.
 
\subsection{Spin temperature and evolution of the cold ISM in galaxies}

It is now well established that the cosmic star formation rate density underwent a rapid acceleration in the early Universe, peaked at $z \approx 1.5 - 2.5$, and then declined by a factor of $10 - 15$ to the present epoch (\citealt{Madau:2014}). Since stars form in self-gravitating clouds of dense molecular gas imbedded within the ISM (\citealt{McKee:2007}), it is expected that strong evolution should also be observed in the cosmological mass density of the molecular gas in galaxies. Observations of the bulk tracers of molecular gas in galaxies -- CO emission (e.g. \citealt{Decarli:2019, Decarli:2020, Lenkic:2020, Riechers:2020a, Riechers:2020b, Fletcher:2021}), supplemented by far-infrared and mm-wavelength observations of the dust continuum (e.g. \citealt{Berta:2013, Scoville:2017, Magnelli:2020}) -- support this expected strong evolution.

In \autoref{figure:tspin_vs_z} we show the best fitting parametric curve to the observational data obtained by \cite{Walter:2020}, alongside that expected from the depletion-time model of \cite{Tacconi:2020}; both agree within the uncertainties of the observational data and show a peaked-evolution that mirrors that of the SFR density (albeit slightly delayed). By balancing the flow rates between phases of the baryonic matter, \cite{Walter:2020} showed that the observed changes in SFR and multi-phase gas densities can be simply modelled by changes in the net conversion rates of ionised to atomic gas, and atomic to molecular gas. Both appear to have declined by an order of magnitude since $z \approx 2$. 

However, when examined in detail the atomic gas in galaxies is far more complex; it is  multi-phased and spans two orders of magnitude in density and temperature (\citealt{Wolfire:2003}). As mentioned already in this paper, recent observations of the Milk Way ISM show the \mbox{H\,{\sc i}} to exist in two stable phases in pressure balance, the denser CNM ($T_{\rm s} \sim 100$\,K) and the more diffuse WNM ($T_{\rm s} \sim 10\,000$\,K), as well as a substantial fraction of unstable UNM ($T_{\rm s} \sim 500$\,K) that is generated by dynamical processes such as turbulence and supernova shocks (\citealt{Heiles:2003, Murray:2018}). If the denser and cooler CNM is a pre-requisite for molecular cloud and star formation (e.g. \citealt{Krumholz:2009}), then evolution of the conversion rate of \mbox{H\,{\sc i}} to H$_2$ should be matched by a similar evolution in the relative fraction of the warm to cold phase atomic gas. 

Evolution in the physical state of the atomic gas would be traced by the \mbox{H\,{\sc i}} spin temperature inferred from 21-cm absorption line surveys. As yet no strong constraints on the detailed evolution of $T_{\rm s}$ are possible from the existing data shown in \autoref{figure:tspin_vs_z}. However, we note that at intermediate cosmological redshifts ($z \sim 0.5 - 1$) both our upper limit on $T_{\rm s}$ and the DLAs in the sample of K14 have values that are consistent with being lower than that of the Milky Way ISM at $z = 0$. This could indicate evolution of the ISM towards higher CNM fractions at intermediate cosmological redshifts, as would be expected for a higher molecular gas density.

\section{Conclusions}

We use a Bayesian technique to infer the harmonic mean spin temperature from a recent 21-cm absorption-line survey with the ASKAP telescope towards 53 compact radio sources (S20). Conditional on the outcome of four detections, we obtain a 95\,per\,cent upper limit on the harmonic mean spin temperature of $T_{\rm s} \leq 274$\,K in DLAs at $0.37 < z < 1.00$. This measurement is consistent with 21-cm absorption-line surveys of known individual DLAs at the same redshift and is possibly indicative of higher fractions of cold phase gas ($T \sim 100$\,K) at $z \sim 1$ than is typical of the local ISM. Such a result would be expected if the cold fraction in galaxies evolves similarly to the molecular and star forming densities, resulting from changes in the physical conditions of the interstellar medium. However, larger 21-cm line surveys are required to verify such a claim.

For the data considered in this work, the total absorption path length sensitive to 21-cm absorption is only $\Delta{X} \sim 10 - 100$. However, future large-scale surveys with the pathfinder telescopes to the Square Kilometre Array, including ASKAP FLASH (e.g. A20, Allison et al. in preparation) and the MeerKAT Absorption Line Survey (MALS; \citealt{Gupta:2016}), are expected to span intervals that are three orders of magnitude larger ($\Delta{X} \sim 10^{4} - 10^{5}$) and will provide correspondingly stronger constraints on the evolution of cold gas in galaxies. By applying the method presented in this paper, we expect these larger surveys to constrain the harmonic mean spin temperature to a precision of $\sim$ 10\,per\,cent (\citealt{Allison:2016}), allowing self-consistent direct measurements of evolution of the cold phase gas in several redshift bins. 

\section*{Acknowledgements}

We are grateful to Robert Allison and Hengxing Pan for useful discussions on Bayesian priors. We also thank Elaine Sadler for helpful comments on a previous version of the manuscript. Finally, we thank the reviewer, Jeremy Darling, for comments that helped improve the presentation and clarity of the paper.

JRA acknowledges support from a Christ Church Career Development Fellowship. Parts of this research were conducted by the Australian Research Council Centre of Excellence for All-sky Astrophysics in 3D (ASTRO 3D) through project number CE170100013. 

This work was supported by resources provided by the Pawsey Supercomputing Centre with funding from the Australian Government and the Government of Western Australia, including computational resources provided by the Australian Government under the National Computational Merit Allocation Scheme (project JA3).

The Australian SKA Pathfinder is part of the Australia Telescope National Facility which is managed by CSIRO. Operation of ASKAP is funded by the Australian Government with support from the National Collaborative Research Infrastructure Strategy. ASKAP uses the resources of the Pawsey Supercomputing Centre. Establishment of ASKAP, the Murchison Radio-astronomy Observatory and the Pawsey Supercomputing Centre are initiatives of the Australian Government, with support from the Government of Western Australia and the Science and Industry Endowment Fund. We acknowledge the Wajarri Yamatji people as the traditional owners of the Observatory site.

We have made use of \textsc{Astropy}, a community-developed core \textsc{PYTHON} package for astronomy (\citealt{Astropy:2013}), and NASA's Astrophysics Data System Bibliographic Services.

%%%%%%%%%%%%%%%%%%%%%%%%%%%%%%%%%%%%%%%%%%%%%%%%%%

\section*{Data Availability}

The data underlying this article were accessed from the Australia Telescope National Facility (ATNF). The derived data generated in this research were originally published by \cite{Sadler:2020} and \cite{Allison:2020}, and will be shared on reasonable request to the corresponding author. 

% The inclusion of a Data Availability Statement is a requirement for articles published in MNRAS. Data Availability Statements provide a standardised format for readers to understand the availability of data underlying the research results described in the article. The statement may refer to original data generated in the course of the study or to third-party data analysed in the article. The statement should describe and provide means of access, where possible, by linking to the data or providing the required accession numbers for the relevant databases or DOIs.



\bibliographystyle{mnras}
\bibliography{spin_temperature_early_science} 

%%%%%%%%%%%%%%%%%%%%%%%%%%%%%%%%%%%%%%%%%%%%%%%%%%

%%%%%%%%%%%%%%%%% APPENDICES %%%%%%%%%%%%%%%%%%%%%

% \appendix

% \section{Some extra material}

% If you want to present additional material which would interrupt the flow of the main paper,
% it can be placed in an Appendix which appears after the list of references.

%%%%%%%%%%%%%%%%%%%%%%%%%%%%%%%%%%%%%%%%%%%%%%%%%%

% Don't change these lines
\bsp	% typesetting comment
\label{lastpage}
\end{document}